\documentclass[12pt]{article}
\usepackage{amssymb}
\begin{document}



\def\a{\alpha}
\def\b{\beta}
\def\d{\delta}
\def\e{\epsilon}
\def\g{\gamma}
\def\h{\mathfrak{h}}
\def\k{\kappa}
\def\l{\lambda}
\def\o{\omega}
\def\p{\wp}
\def\r{\rho}
\def\t{\theta}
\def\s{\sigma}
\def\z{\zeta}
\def\x{\xi}
 \def\A{{\cal{A}}}
 \def\B{{\cal{B}}}
 \def\C{{\cal{C}}}
 \def\D{{\cal{D}}}
\def\G{\Gamma}
\def\K{{\cal{K}}}
\def\O{\Omega}
\def\L{\Lambda}
\def\f{E_{\tau,\eta}(sl_2)}
\def\E{E_{\tau,\eta}(sl_n)}
\def\Zb{\mathbb{Z}}
\def\Cb{\mathbb{C}}

\def\R{\overline{R}}

\def\beq{\begin{equation}}
\def\eeq{\end{equation}}
\def\bea{\begin{eqnarray}}
\def\eea{\end{eqnarray}}
\def\ba{\begin{array}}
\def\ea{\end{array}}
\def\no{\nonumber}
\def\le{\langle}
\def\re{\rangle}
\def\lt{\left}
\def\rt{\right}

\newtheorem{Theorem}{Theorem}
\newtheorem{Definition}{Definition}
\newtheorem{Proposition}{Proposition}
\newtheorem{Lemma}{Lemma}
\newtheorem{Corollary}{Corollary}
\newcommand{\proof}[1]{{\bf Proof. }
        #1\begin{flushright}$\Box$\end{flushright}}

\baselineskip=20pt

\newfont{\elevenmib}{cmmib10 scaled\magstep1}
\newcommand{\preprint}{
   \begin{flushleft}
     \elevenmib Yukawa\, Institute\, Kyoto\\
   \end{flushleft}\vspace{-1.3cm}
   \begin{flushright}\normalsize  \sf
     YITP-03-53\\
     {\tt hep-th/0308118} \\ August 2003
   \end{flushright}}
\newcommand{\Title}[1]{{\baselineskip=26pt
   \begin{center} \Large \bf #1 \\ \ \\ \end{center}}}
\newcommand{\Author}{\begin{center}
   \large \bf
W.\,-L. Yang$,{}^{a,b,}$\footnote{Address after May 2004:
Department  of Mathematics, The University of Queensland, Brisbane
4072, Australia.}
 ~ and~R.~Sasaki${}^b$\end{center}}
\newcommand{\Address}{\begin{center}

     ${}^a$ Institute of Modern Physics, Northwest University\\
     Xian 710069, P.R. China\\
     ~~\\
     ${}^b$ Yukawa Institute for Theoretical Physics,\\
     Kyoto University, Kyoto 606-8502, Japan
   \end{center}}
\newcommand{\Accepted}[1]{\begin{center}
   {\large \sf #1}\\ \vspace{1mm}{\small \sf Accepted for Publication}
   \end{center}}

\preprint
\thispagestyle{empty}
\bigskip\bigskip\bigskip

\Title{Solution of the dual reflection equation for
$A^{(1)}_{n-1}$ SOS  model } \Author

\Address
\vspace{1cm}

\begin{abstract}
We obtain a diagonal solution of the {\it dual reflection}
equation for the  elliptic $A^{(1)}_{n-1}$ SOS model. The
isomorphism between the solutions of the reflection equation and
its dual is studied.

\vspace{1truecm}
\end{abstract}
\newpage
\section{Introduction}
\label{intro} \setcounter{equation}{0}

Two-dimensional lattice spin models in statistical mechanics have
traditionally been solved by imposing periodic boundary condition.
The Yang-Baxter equation \cite{Yan67,Bax82}\bea
R_{12}(u_1-u_2)R_{13}(u_1-u_3)R_{23}(u_2-u_3)=
R_{23}(u_2-u_3)R_{13}(u_1-u_3)R_{12}(u_1-u_2), \label{YBE-V} \eea
together with such boundary condition then leads to families of
commuting row {\it transfer matrices} and hence solvability
\cite{Bax82}. The work of Sklyanin \cite{Skl88} shows that, by
using the reflection equation (RE) introduced by Cherednik
\cite{Che84}
 \bea R_{12}(u_1-u_2)K_1(u_1)R_{21}(u_1+u_2)K_2(u_2)=
K_2(u_2)R_{12}(u_1+u_2)K_1(u_1)R_{21}(u_1-u_2),\label{RE-V}\eea it
is also possible to construct families of commuting {\it
double-row transfer matrices} for vertex models with open boundary
conditions. Then such a scheme has been generalized to {\it face
type} solid-on-solid (SOS) models \cite{Beh96, Fan97}.

In order to construct the {\it double-row transfer matrices},
besides the RE, one  needs the dual reflection equation whose
explicit form is related with the crossing-unitarity relation of
the R-matrix \cite{Skl88,Mez91,Beh96,Fan97}. For the $\Zb_n$
Belavin model \cite{Bel81}, the dual RE reads \cite{Fan97} \bea
&&R_{12}(u_2-u_1)\tilde{K}_1(u_1)R_{21}(-u_1-u_2-nw)\tilde{K}_2(u_2)\no\\
&&~~~~~~=
\tilde{K}_2(u_2)R_{12}(-u_1-u_2-nw)\tilde{K}_1(u_1)R_{21}(u_2-u_1),
\label{DRE-V}\eea where $w$ is the crossing parameter of the
R-matrix. Moreover, there exists a {\it simple-form\/} isomorphism
between the solution of the RE (\ref{RE-V}) and that of its dual
(\ref{DRE-V}) \bea
\tilde{K}(u)=K(-u-\frac{nw}{2}).\label{ISO-V}\eea However, for
  integrable SOS models, due to the complicated
crossing-unitarity  relation of R-matrix (Boltzmann weight)
(\ref{Crossing-F}) \cite{Jim88,Kun91}, the dual RE (\ref{DRE-F})
contains  the face type parameters $\lt\{\l_j\rt\}$ in addition to
the spectral parameter. A generalized isomorphism between the
solutions to the RE and its dual for SOS models, if exists,  is
yet to be found. In this sense, the dual RE for the face type
models has got its own {\it independent\/} role in contrast with
the vertex model.

The RE of SOS models has been solved to give the diagonal
K-matrices for the $A^{(1)}_n,~B^{(1)}_n,$ $C^{(1)}_n,~D^{(1)}_n$,
$A^{(2)}_{2n}$ and $A^{(2)}_{2n+1}$ SOS models \cite{Bat96}. But
the generic (non-diagonal) K-matrix is known only for the
$A^{(1)}_1$ SOS model \cite{Fan95,Beh96}. However, the dual RE of
the face type was solved only for the $A^{(1)}_1$ SOS model
\cite{Beh96}. In this paper, we consider the dual RE for the
$A^{(1)}_{n-1}$ SOS model. After briefly reviewing the face-vertex
correspondence between the $\Zb_n$ Belavin model and the
$A^{(1)}_{n-1}$ SOS model \cite{Jim87}, we construct the
isomorphism between the solution of the RE and its dual for the
$A^{(1)}_{n-1}$ SOS model in section 3. In section 4, we derive a
diagonal solution to the dual RE by solving directly. Then we
prove that our diagonal solution to the dual RE can be obtained
through  the isomorphism transformation (\ref{ISO-F}) from the
diagonal solution \cite{Bat96} of RE by a special choice of the
free parameter $\l'$. The final section is for conclusions.

\section{Reflection equation and its dual for
 $A^{(1)}_{n-1}$ SOS model}
 \label{BAE} \setcounter{equation}{0}

\subsection{ $\Zb_n$ Belavin R-matrix}

Let us fix $\tau$ such that $Im(\tau)>0$ and a generic complex
number $w$. Introduce the following elliptic functions \bea
&&\t\lt[
\begin{array}{c}
a\\b
\end{array}\rt](u,\tau)=\sum_{m=-\infty}^{\infty}
exp\lt\{\sqrt{-1}\pi\lt[(m+a)^2\tau+2(m+a)(u+b)\rt]\rt\},\\
&&\t^{(j)}(u)=\t\lt[\begin{array}{c}\frac{1}{2}-\frac{j}{n}\\[2pt]\frac{1}{2}
\end{array}\rt](u,n\tau),~~~
\s(u)=\t\lt[\begin{array}{c}\frac{1}{2}\\[2pt]\frac{1}{2}
\end{array}\rt](u,\tau).\label{Function}
\eea Among them the $\s$-function\footnote{Our $\s$-function is
the $\vartheta$-function $\vartheta_1(u)$ \cite{Whi50}. It has the
following relation with the {\it Weierstrassian\/} $\s$-function
if denoted it by $\s_w(u)$: $\s_w(u)\propto e^{\eta_1u^2}\s(u)$,
$\eta_1=\pi^2(\frac{1}{6}-4\sum_{n=1}^{\infty}\frac{nq^{2n}}{1-q^{2n}})
$ and $q=e^{\sqrt{-1}\tau}$.} satisfies the following
identity:\bea
&&\s(u+x)\s(u-x)\s(v+y)\s(v-y)-\s(u+y)\s(u-y)\s(v+x)\s(v-x)\no\\
&&~~~~~~=\s(u+v)\s(u-v)\s(x+y)\s(x-y),\label{identity}\eea which
will be useful in the following.

Let $R^B(u)\in End(\Cb^n\otimes\Cb^n)$ be the $\Zb_n$ Belavin
R-matrix \cite{Bel81} given by  \bea
R^B(u)=\sum_{i,j,k,l}R^{kl}_{ij}(u)E_{ik}\otimes
E_{lj},\label{Belavin-R}\eea in which $E_{ij}$ is the matrix with
elements $(E_{ij})^l_k=\d_{jk}\d_{il}$. The coefficient functions
are \cite{Jim88} \bea R^{kl}_{ij}(u)=\lt\{
\begin{array}{ll}
\frac{h(u)\s(w)\t^{(i-j)}(u+w)}{\s(u+w)\t^{(i-k)}(w)\t^{(k-j)}(u)}&{\rm
if}~i+j=k+l~mod~n,\\[6pt]
0&{\rm otherwise.}
\end{array}\rt.
\eea Here we have set \bea
h(u)=\frac{\prod_{j=0}^{n-1}\t^{(j)}(u)}
{\prod_{j=1}^{n-1}\t^{(j)}(0)}.\eea The R-matrix satisfies the
quantum Yang-Baxter equation (\ref{YBE-V}) and the following
unitarity and crossing-unitarity relations \cite{Ric86}\bea
&&\hspace{-1.5cm}\mbox{
Unitarity}:\,{R^B}_{12}(u)R^B_{21}(-u)= id,\label{Unitarity}\\
&&\hspace{-1.5cm}\mbox{
Crossing-unitarity}:\,(R^B)^{t_2}_{21}(-u-nw)(R^B)_{12}^{t_2}(u)
=\frac{e^{\sqrt{-1}nw}\s(u)\s(u+nw)}{\s(u+w)\s(u+nw-w)}~id,
\label{crosing-unitarity} \eea where $t_i$ denotes the
transposition in the $i$-th space.
\subsection{$A^{(1)}_{n-1}$ SOS R-matrix and face-vertex correspondence}

Let $\lt\{\e_{i}~|~i=1,2,\cdots,n\rt\}$ be the orthonormal basis
of the vector space $\Cb^n$ such that $\langle\e_i,~\e_j
\rangle=\d_{ij}$. The $A_{n-1}$ simple roots are
$\lt\{\a_{i}=\e_i-\e_{i+1}~|~i=1,\cdots,n-1\rt\}$ and the
fundamental weights $\lt\{\L_i~|~i=1,\cdots,n-1\rt\}$ satisfying
$\langle\L_i,~\a_j\rangle=\d_{ij}$ are given by \bea
\L_i=\sum_{k=1}^{i}\e_k-\frac{i}{n}\sum_{k=1}^{n}\e_k. \no\eea Set
\bea \hat{\imath}=\e_i-\overline{\e},~~\overline{\e}=
\frac{1}{n}\sum_{k=1}^{n}\e_k,~~i=1,\cdots,n,~~{\rm
then}~~\sum_{i=1}^n\hat{\imath}=0. \label{Vectors} \eea For each
dominant weight $\L=\sum_{i=1}^{n-1}a_i\L_{i}~,~~a_{i}\in \Zb^+$,
there exists an irreducible highest weight finite-dimensional
representation $V_{\L}$ of $A_{n-1}$ with the highest vector $
|\L\rangle$. For example the fundamental vector representation is
$V_{\L_1}$.

Let $\h$ be the Cartan subalgebra of $A_{n-1}$ and $\h^{*}$ be its
dual. A finite dimensional diagonalizable  $\h$-module is a
complex finite dimensional vector space $W$ with a weight
decomposition $W=\oplus_{\mu\in \h^*}W[\mu]$, so that $\h$ acts on
$W[\mu]$ by $x\,v=\mu(x)\,v$, $(x\in \h,~~v\in~W[\mu])$. For
example, the fundamental vector representation $V_{\L_1}=\Cb^n$,
the non-zero weight spaces $W[\hat{\imath}]=\Cb
\e_i,~i=1,\cdots,n$.

For a generic $\l\in \Cb^n$, define \bea
\l_i=\langle\l,\e_i\rangle,
~~\l_{ij}=\l_i-\l_j,~~|\l|=\sum_{l=1}^n\l_l,~~i,j=1,\cdots,n.
\label{Def1}\eea Let $R(z,\l)\in End(\Cb^n\otimes\Cb^n)$ be the
R-matrix of the $A^{(1)}_{n-1}$ SOS model given by \bea
\hspace{-0.5cm}R(z,\l)=\sum_{i=1}^{n}R^{ii}_{ii}(z,\l)E_{ii}\otimes
E_{ii} +\sum_{i\ne j}\lt\{R^{ij}_{ij}(z,\l)E_{ii}\otimes E_{jj}+
R^{ji}_{ij}(z,\l)E_{ji}\otimes E_{ij}\rt\}.\label{R-matrix} \eea
The coefficient functions are \bea &&R^{ii}_{ii}(z,\l)=1,~~
R^{ij}_{ij}(z,\l)=\frac{\s(z)\s(\l_{ij}w-w)}
{\s(z+w)\s(\l_{ij}w)},\label{Elements1}\\
&& R^{ji}_{ij}(z,\l)=\frac{\s(w)\s(z+\l_{ij}w)}
{\s(z+w)\s(\l_{ij}w)},\label{Elements2}\eea  and  $\l_{ij}$ is
defined in (\ref{Def1}). The R-matrix satisfies the dynamical
(modified) quantum Yang-Baxter equation \bea
&&R_{12}(z_1-z_2,\l-h^{(3)})R_{13}(z_1-z_3,\l)
R_{23}(z_2-z_3,\l-h^{(1)})\no\\
&&~~~~=R_{23}(z_2-z_3,\l)R_{13}(z_1-z_3,\l-h^{(2)})R_{12}(z_1-z_2,\l),
\label{MYBE}\eea with unitarity  relation \bea
R_{12}(u,\l)R_{21}(-u,\l)=id.\label{Initial}\eea We adopt the
notation: $R_{12}(z,\l-h^{(3)})$ acts on a tensor $v_1\otimes v_2
\otimes v_3$ as $R(z,\l-\mu)\otimes id$ if $v_3\in W[\mu]$. Let us
introduce \bea &&\tilde{R}(u,\l)_{ij}^{kl}=R(u,\l)^{kl}_{ij}\lt\{
\frac{f_2(\l;k)}{f_2(\l+\hat{k};k)}
\frac{f_2(\l+\hat{\imath}+\hat{\jmath};i)}{f_2(\l+\hat{\jmath};i)}\rt\},\label{TR}\\
&&f_2(\l;j)=\prod_{k\ne j}\frac{\s(\l_{jk}w)}{\s(w)}.
\label{F-function} \eea The R-matrix satisfies the following
crossing-unitarity relation \cite{Fan97}\bea
\sum_{i_2,j_2=1}^{n}\tilde{R}(-u-nw,\l-\hat{\jmath}_2)^{j_2~i_1}_{j_1~i_2}
R(u,\l-\hat{\jmath}_2)^{i_2~j_3}_{i_3~j_2}=
\frac{e^{\sqrt{-1}nw}\s(u)\s(u+nw)}{\s(u+w)\s(u+nw-w)}~
\d^{i_1}_{i_3}\d^{j_1}_{j_3}.\label{Crossing-F}\eea

Let us introduce an intertwiner---an $n$-component column vector
$\phi_{\l,\l-\hat{\jmath}}(u)$ whose $k$-th element is \bea
\phi^{(k)}_{\l,\l-\hat{\jmath}}(u)=\t^{(k)}(u+nw\l_j).\label{Intertw-d}\eea
Using the intertwiner, the face-vertex correspondence can be
written as \cite{Jim87}\bea R^B_{12}(u_1-u_2)
\phi_{\l,\l-\hat{\imath}}(u_1)\hspace{-0.1cm}\otimes
\hspace{-0.1cm}
\phi_{\l-\hat{\imath},\l-\hat{\imath}-\hat{\jmath}}(u_2)
\hspace{-0.1cm}=
\hspace{-0.2cm}\sum_{kl}\hspace{-0.1cm}R(u_1-u_2,\l)^{kl}_{ij}
\phi_{\l-\hat{l},\l-\hat{l}-\hat{k}}(u_1)\hspace{-0.1cm} \otimes
\hspace{-0.1cm}\phi_{\l,\l-\hat{l}}(u_2).\label{Face-vertex}\eea
Then the Yang-Baxter equation of the $\Zb_n$ Belavin R-matrix
$R^B(u)$ (\ref{YBE-V}) is equivalent to the dynamical Yang-Baxter
equation of the $A^{(1)}_{n-1}$ SOS R-matrix $R(u,\l)$
(\ref{MYBE}).

\section{RE and dual RE for $A^{(1)}_{n-1}$ SOS model}
\label{RE} \setcounter{equation}{0}

In this section, using the intertwiner between the $\Zb_n$ Belavin
R-matrix and that of  the $A^{(1)}_{n-1}$ SOS model, we construct
the isomorphism between the solution of the RE for the
$A^{(1)}_{n-1}$ SOS model and that of its dual from the
isomorphism (\ref{ISO-V}).
\subsection{RE and its dual for  SOS model}
The RE of the K-matrix $\K(\l|u)$ for the face type SOS  model was
given as follows \cite{Beh96,Fan95,Kul95,Ahn95} \bea
&&\sum_{i_1,i_2}\sum_{j_1,j_2}~
R(u_1-u_2,\l)^{i_0\,j_0}_{i_1\,j_1}\K(\l+\hat{\jmath}_1+\hat{\imath}_2|u_1)
^{i_1}_{i_2}R(u_1+u_2,\l)^{j_1\,i_2}_{j_2\,i_3}
\K(\l+\hat{\jmath}_3+\hat{\imath}_3|u_2)^{j_2}_{j_3}\no\\
&&~~~~=\sum_{i_1,i_2}\sum_{j_1,j_2}~
\K(\l+\hat{\jmath}_1+\hat{\imath}_0|u_2)
^{j_0}_{j_1}R(u_1+u_2,\l)^{i_0\,j_1}_{i_1\,j_2}
\K(\l+\hat{\jmath}_2+\hat{\imath}_2|u_1)^{i_1}_{i_2}
R(u_1-u_2,\l)^{j_2\,i_2}_{j_3\,i_3}.\no\\
&&~~\label{RE-F} \eea The dual RE of the K-matrix
$\tilde{\K}(\l|u)$ was written down by \cite{Beh96,Fan97} \bea
&&\sum_{i_1,i_2}\sum_{j_1,j_2}~
R(u_2-u_1,\l)^{i_0\,j_0}_{i_1\,j_1}\tilde{\K}(\l+\hat{\jmath}_1+\hat{\imath}_1|u_1)
^{i_1}_{i_2}\tilde{R}(-u_1-u_2-nw,\l)^{j_1\,i_2}_{j_2\,i_3}
\tilde{\K}(\l+\hat{\jmath}_2+\hat{\imath}_3|u_2)^{j_2}_{j_3}\no\\
&&~=\sum_{i_1,i_2}\sum_{j_1,j_2}~
\tilde{\K}(\l+\hat{\jmath}_0+\hat{\imath}_0|u_2)
^{j_0}_{j_1}\tilde{R}(-u_1-u_2-nw,\l)^{i_0\,j_1}_{i_1\,j_2}
\tilde{\K}(\l+\hat{\jmath}_2+\hat{\imath}_1|u_1)^{i_1}_{i_2}
R(u_2-u_1,\l)^{j_2\,i_2}_{j_3\,i_3},\no\\
&&~~\label{DRE-F} \eea where $\tilde{R}(u,\l)$ is defined in
(\ref{TR}) for the $A^{(1)}_{n-1}$ SOS model. The explicit
expressions of $\tilde{R}(u,\l)$ for other types of  SOS models
were given in \cite{Fan97}. Because of  the {\it non-trivial}
dependence on the face type parameters $\{\l_j\}$, the dual RE of
SOS models should be treated separately in contrast with those of
the vertex models.

As in the Sklyanin scheme for the vertex models, one can construct
families of commuting {\it double-row transfer matrices} for the
SOS model with open boundary condition in terms of the K-matrices
$\K(\l|u)$ and $\tilde{\K}(\l|u)$ \cite{Beh96,Fan97}.

\subsection{Isomorphism between the solutions of the RE and
its dual for $A^{(1)}_{n-1}$ SOS model}

Thanks to the face-vertex correspondence between the  $\Zb_n$
Belavin vertex model and the $A^{(1)}_{n-1}$ SOS model
(\ref{Face-vertex}), we can construct the isomorphism between the
solutions of the RE and its dual for the $A^{(1)}_{n-1}$ SOS model
from the isomorphism (\ref{ISO-V}) of  the $\Zb_n$ Belavin vertex
model.

Let us introduce other types of  intertwiners $\bar{\phi}$ and
$\tilde{\phi}$ satisfying the following orthogonality conditions
\bea &&\sum_{k}\bar{\phi}^{(k)}_{\l,\l-\hat{\imath}}(u)
~\phi^{(k)}_{\l,\l-\hat{\jmath}}(u)=\d_{ij},\label{Int1}\\
&&\sum_{k}\tilde{\phi}^{(k)}_{\l+\hat{\imath},\l}(u)
~\phi^{(k)}_{\l+\hat{\jmath},\l}(u)=\d_{ij}.\label{Int2}\eea One
can derive the ``completeness" relations from  the above
conditions\bea &&\sum_{k}\bar{\phi}^{(i)}_{\l,\l-\hat{k}}(u)
~\phi^{(j)}_{\l,\l-\hat{k}}(u)=\d_{ij},\label{Int3}\\
&&\sum_{k}\tilde{\phi}^{(i)}_{\l+\hat{k},\l}(u)
~\phi^{(j)}_{\l+\hat{k},\l}(u)=\d_{ij},\label{Int4}\eea and the
following relation between the intertwiners $\bar{\phi}$ and
$\tilde{\phi}$ from their definitions (\ref{Int1}) and
(\ref{Int2}) \cite{Fan97} \bea
\bar{\phi}_{\l+\hat{\jmath},\l}(u)=\frac{\s(u+w|\l|-\frac{n-1}{2}-w)}
{\s(u+w|\l|-\frac{n-1}{2})}\lt\{\prod_{k\neq
j}\frac{\s(\l_{jk}w)}{\s(\l_{jk}w+w)}\rt\}\tilde{\phi}
_{\l+\hat{\jmath},\l}(u-nw).\label{relation-1}\eea Noting the fact
$\langle \bar{\e},\e_j\rangle=\frac{1}{n}$ and the definition of
the intertwiner (\ref{Intertw-d}), one can derive the following
relations: for $\forall \a\in\Cb$ \bea
&&\phi_{\l+\a\bar{\e},\l+\a\bar{\e}-\hat{\jmath}}(u)=
\phi_{\l,\l-\hat{\jmath}}(u+\a w),\label{relat-1-1}\\
&&\bar{\phi}_{\l+\a\bar{\e},\l+\a\bar{\e}-\hat{\jmath}}(u)=
\bar{\phi}_{\l,\l-\hat{\jmath}}(u+\a w),\\
&&\tilde{\phi}_{\l+\a\bar{\e},\l+\a\bar{\e}-\hat{\jmath}}(u)=
\tilde{\phi}_{\l,\l-\hat{\jmath}}(u+\a w). \label{relation-3}\eea

Define \bea &&\K(\l|u)^j_i=\sum_{s,t}\tilde{\phi}^{(s)}
_{\l-\hat{\imath}+\hat{\jmath},~\l-\hat{\imath}}(u)K(u)^s_t\phi^{(t)}
_{\l,~\l-\hat{\imath}}(-u),\label{F-V1}\\
&&\tilde{\K}(\l|u)^j_i=\sum_{s,t}\bar{\phi}^{(s)}
_{\l,~\l-\hat{\jmath}}(-u)\tilde{K}(u)^s_t\phi^{(t)}
_{\l-\hat{\jmath}+\hat{\imath},~\l-\hat{\jmath}}(u).\label{F-V2}
\eea Then we have \begin{Theorem} \cite{Fan97} The above relations
(\ref{F-V1}), (\ref{F-V2}) map the solutions $K(u)$ and
$\tilde{K}(u)$ to the RE (\ref{RE-V}) and the dual (\ref{DRE-V})
for the $\Zb_n$ Belavin R-matrix to the solutions $\K(\l|u)$ and
$\tilde{\K}(\l|u)$ to  the RE (\ref{RE-F}) and the dual
(\ref{DRE-F}) for  the $A^{(1)}_{n-1}$ SOS R-matrix, and vice
versa.
\end{Theorem}
\vspace{0.6truecm}

Using the relations (\ref{Int3}) and (\ref{Int4}), one can invert
(\ref{F-V1}) \bea K(u)^s_t=
\sum_{i,j}\phi^{(s)}_{\l-\hat{\imath}+\hat{\jmath},~\l-\hat{\imath}}(u)
\K(\l|u)^j_i\bar{\phi}^{(t)}_{\l,~\l-\hat{\imath}}(-u).\eea Using
the isomorphism (\ref{ISO-V}) between the solutions of the RE and
the dual RE for the $\Zb_n$ Belavin R-matrix, the relations
(\ref{Int3}), (\ref{Int4}) and (\ref{F-V2}), we have \bea
&&\tilde{\K}(\l|u)^{\nu}_{\mu}=\sum_{s,t} \bar{\phi}^{(s)}
_{\l,~\l-\hat{\nu}}(-u)\tilde{K}(u)^s_t\phi^{(t)}
_{\l-\hat{\nu}+\hat{\mu},~\l-\hat{\nu}}(u)\no\\
&&~~~~~~=\sum_{s,t}\bar{\phi}^{(s)}
_{\l,~\l-\hat{\nu}}(-u)K(-u-\frac{nw}{2})^s_t
\phi^{(t)}_{\l-\hat{\nu}+\hat{\mu},~\l-\hat{\nu}}(u)\no\\
&&~~~~~~=\sum_{i,j}\sum_{s,t}\bar{\phi}^{(s)}
_{\l,~\l-\hat{\nu}}(-u)
\phi^{(s)}_{\l'-\hat{\imath}+\hat{\jmath},~\l'-\hat{\imath}}
(-u-\frac{nw}{2})
\K(\l'|-u-\frac{nw}{2})^j_i\no\\
&&~~~~~~~~~~~~~~~~\times
\bar{\phi}^{(t)}_{\l',~\l'-\hat{\imath}}(u+\frac{nw}{2})
\phi^{(t)}_{\l-\hat{\nu}+\hat{\mu},~\l-\hat{\nu}}(u) \no\\
&&~~~~~~=\sum_{i,j}M(\l,~\l'-\hat{\imath}|-u)^{\nu}_j
\K(\l'|-u-\frac{nw}{2})^j_iM(\l',~\l-\hat{\nu}|u+\frac{nw}{2})^i_{\mu},
\eea where $\l'\in \Cb^n$ is arbitrary  and a {\it crossing
matrix\/} $M(\l,~\l'|u)^{\nu}_j$ is defined  by\bea
M(\l,~\l'|u)^{\nu}_j=
\sum_{t}\bar{\phi}^{(t)}_{\l,~\l-\hat{\nu}}(u)
\phi^{(t)}_{\l'+\hat{\jmath},~\l'}(u-\frac{nw}{2}).\label{cross-matrix}\eea
Finally, we obtain
\begin{Theorem}
The solutions to the RE (\ref{RE-F}) and the dual (\ref{DRE-F})
for the $A^{(1)}_{n-1}$ SOS R-matrix have the following
isomorphism \bea
&&\tilde{\K}(\l|u)^{\nu}_{\mu}=\sum_{i,j}M(\l,~\l'-\hat{\imath}|-u)^{\nu}_j
\K(\l'|-u-\frac{nw}{2})^j_iM(\l',~\l-\hat{\nu}|u+\frac{nw}{2})^i_{\mu},
\label{ISO-F}\eea where   $\l'\in \Cb^n$ is arbitrary.
\end{Theorem}
\vspace{0.6truecm}

We remark that the crossing matrix (\ref{cross-matrix}) is
generally {\it non-diagonal\/}. Hence, the corresponding
$\tilde{\K}(\l|u)$ of the solution to the dual RE (\ref{DRE-F})
obtained by the isomorphism (\ref{ISO-F}) from the diagonal
solution \cite{Bat96} to RE is generally non-diagonal, too, except
for the case that  a special choice of ``{\it moduli\/}" parameter
$\l'$ is chosen as (\ref{relation-2}) (this special case will be
clarified later in the next section). However, in order to
diagonalize the corresponding {\it double-row transfer matrices\/}
for the $A^{(1)}_{n-1}$ SOS model by the algebraic Bethe ansatz
method, one needs $\K(\l|u)$ and $\tilde{\K}(\l|u)$ both diagonal
\cite{Fan96,Sas03}. In the next section, we shall search for a
diagonal $\tilde{\K}(\l|u)$.

\section{Diagonal solution of the dual RE for $A^{(1)}_{n-1}$ SOS
model} \label{DRE} \setcounter{equation}{0}

In this section we look for the diagonal solution to the dual RE
(\ref{DRE-F}) for  the $A^{(1)}_{n-1}$ SOS model, namely, the
K-matrix $\tilde{\K}(\l|u)$ of following form \bea
\tilde{\K}(\l|u)^j_i=\tilde{k}(\l|u)_i\d^j_i,\label{Form-1}\eea
where $\lt\{\tilde{k}(\l|u)_i\rt\}$ are the functions of the face
parameters $\lt\{\l_j\rt\}$ and the spectral parameter $u$. From
directly solving the equation (\ref{DRE-F}), we have
\begin{Theorem}
For \bea \tilde{k}(\l|u)_i=\lt\{\prod_{k\ne
i}\frac{\s(\l_{ik}w-w)}{\s(\l_{ik}w)}\rt\}
~\frac{\s(\l_iw+\bar{\xi}+u+\frac{nw}{2})}
{\s(\l_iw+\bar{\xi}-u-\frac{nw}{2})} ~f(u,\l),\label{Solution}\eea
in which $\bar{\xi}$ is a free parameter and $f(u,\l)$ is any
non-vanishing  function of $\l$ and $u$, the diagonal K-matrix
$\tilde{\K}(\l|u)$ with entries (\ref{Form-1}) and
(\ref{Solution}) is a solution to the dual RE (\ref{DRE-F}) for
the $A^{(1)}_{n-1}$ SOS model.
\end{Theorem}
\vspace{0.6truecm}

\noindent {\it \bf Proof.} Substituting $\tilde{\K}(\l|u)$ of form
(\ref{Form-1}) into the dual RE (\ref{DRE-F}) for the
$A^{(1)}_{n-1}$ SOS model, one finds the only nontrivial
conditions of $\tilde{k}(\l|u)_i$ are \bea
&&R(u_2-u_1,\l)^{ji}_{ji}
\tilde{k}(\l+\hat{\imath}+\hat{\jmath}|u_1)_j~\tilde{R}(-u_1-u_2-nw,\l)^{ij}_{ji}
\tilde{k}(\l+\hat{\imath}+\hat{\jmath}|u_2)_j\no\\
&&~~~~~~~~ +R(u_2-u_1,\l)^{ji}_{ij}
\tilde{k}(\l+\hat{\imath}+\hat{\jmath}|u_1)_i~\tilde{R}(-u_1-u_2-nw,\l)^{ji}_{ji}
\tilde{k}(\l+\hat{\imath}+\hat{\jmath}|u_2)_j\no\\
&&~~=R(u_2-u_1,\l)^{ji}_{ji}
\tilde{k}(\l+\hat{\imath}+\hat{\jmath}|u_1)_i~\tilde{R}(-u_1-u_2-nw,\l)^{ji}_{ij}
\tilde{k}(\l+\hat{\imath}+\hat{\jmath}|u_2)_i\no\\
&&~~~~~~~~ +R(u_2-u_1,\l)^{ij}_{ji}
\tilde{k}(\l+\hat{\imath}+\hat{\jmath}|u_1)_j~\tilde{R}(-u_1-u_2-nw,\l)^{ji}_{ji}
\tilde{k}(\l+\hat{\imath}+\hat{\jmath}|u_2)_i ,~~i\ne j.\no\eea
Substituting (\ref{TR}) and (\ref{Solution}) into the above
equation, the dual RE (\ref{DRE-F}) is equivalent to the following
equation \bea &&\lt\{
\s(u_{-}+\l_{ij}w)\s(u_{+})-\s(u_{-})\s(u_{+}-\l_{ij}w)
\frac{\s(\l_jw+\bar{\xi}'-u'_1)\s(\l_iw+\bar{\xi}'+u'_1)}
{\s(\l_jw+\bar{\xi}'+u'_1)\s(\l_iw+\bar{\xi}'-u'_1)}\rt\}\no\\
&&~~~~~~~~~~~~~~\times
\frac{\s(\l_jw+\bar{\xi}'-u'_2)\s(\l_iw+\bar{\xi}'+u'_2)}
{\s(\l_jw+\bar{\xi}'+u'_2)\s(\l_iw+\bar{\xi}'-u'_2)}\no\\
&&~~=\s(u_{+}+\l_{ij}w)\s(u_{-})-\s(u_{+})\s(u_{-}-\l_{ij}w)
\frac{\s(\l_jw+\bar{\xi}'-u'_1)\s(\l_iw+\bar{\xi}'+u'_1)}
{\s(\l_jw+\bar{\xi}'+u'_1)\s(\l_iw+\bar{\xi}'-u'_1)},\no\\
&&~~\label{Solution1}\eea where $u_{-}=u'_1-u'_2$,
$u_{+}=u'_1+u'_2$, $u'_i=-u_i-\frac{nw}{2}$,
$\bar{\xi}'=\bar{\xi}+\frac{n-2}{n}w$. The equation
(\ref{Solution1}) is a consequence of the  identity
(\ref{identity}). Then we complete our proof.

\vspace{0.6truecm}


Now we shall study the relation between our solution of the dual
RE and the diagonal solution  of RE which was given as follows
\cite{Bat96}\bea \K(\l|u)^j_i=k(\l|u)_i\d^j_i=
g(u,\l)\frac{\s(\l_iw+\xi-u)}{\s(\l_iw+\xi+u)}\d^j_i.\label{Solution-RE}\eea
Here, $g(\l|u)$ is any non-vanishing function of $\l$ and $u$, and
$\xi$ is a free parameter. Let us choose \bea
\l'=\l+\frac{n}{2}\bar{\e}~\Rightarrow~\l'_i=\l_i+\frac{1}{2},
\label{relation-2}\eea the vector $\bar{\e}$ is defined in
(\ref{Vectors}). Using the relation (\ref{relat-1-1}), the
crossing matrix
$M(\l,\l+\frac{n}{2}\bar{\e}-\hat{\imath}|u)^{\nu}_i $ defined in
(\ref{cross-matrix}) becomes simple \bea
&&M(\l,\l+\frac{n}{2}\bar{\e}-\hat{\imath}|u)^{\nu}_i=\sum_{t}
\bar{\phi}^{(t)}_{\l,\l-\hat{\nu}}(u)\phi^{(t)}_{\l+\frac{n}{2}\bar{\e},
\l+\frac{n}{2}\bar{\e}-\hat{\nu}}(u-\frac{nw}{2})\no\\
&&~~~~~~=\sum_{t}
\bar{\phi}^{(t)}_{\l,\l-\hat{\imath}}(u)\phi^{(t)}_{\l,
\l-\hat{\imath}}(u)=\d^{\nu}_i.\eea The resulting solution to the
dual RE by the isomorphism transformation (\ref{ISO-F}) from the
diagonal solution to RE is \bea
\tilde{\K}(\l|u)^{\nu}_{\mu}=k(\l+\frac{n}{2}\bar{\e}|-u-\frac{nw}{2})_{\nu}
M(\l+\frac{n}{2}\bar{\e},
\l-\hat{\nu}|u+\frac{nw}{2})^{\nu}_{\mu}.\eea The relations
(\ref{relation-1}) and (\ref{relation-2}) enable us to further
simplify the expression of the crossing matrix
$M(\l+\frac{n}{2}\bar{\e},
\l-\hat{\nu}|u+\frac{nw}{2})^{\nu}_{\mu}$ \bea
&&M(\l+\frac{n}{2}\bar{\e},
\l-\hat{\nu}|u+\frac{nw}{2})^{\nu}_{\mu}=\frac{\s(u+|\l-\hat{\nu}|
w+\frac{n-2}{2}w-\frac{n-1}{2})}{\s(u+|\l-\hat{\nu}|
w+\frac{n}{2}w-\frac{n-1}{2})} \lt\{
\prod_{k\neq\nu}\frac{\s(\l_{\nu k}w-w)} {\s(\l_{\nu
k}w)}\rt\}\no\\
&&~~~~~~~~~~~~~~~~\times \sum_{t}\tilde{\phi}^{(t)}_
{\l+\frac{n}{2}\bar{\e},\l+\frac{n}{2}\bar{\e}-\hat{\nu}}(u-\frac{nw}{2})
\phi^{(t)}_{\l-\hat{\nu}+\hat{\mu}, \l-\hat{\nu}}(u)\no\\
&&~~~~~~=\frac{\s(u+|\l-\hat{\nu}|
w+\frac{n-2}{2}w-\frac{n-1}{2})}{\s(u+|\l-\hat{\nu}|
w+\frac{n}{2}w-\frac{n-1}{2})} \lt\{
\prod_{k\neq\nu}\frac{\s(\l_{\nu k}w-w)} {\s(\l_{\nu
k}w)}\rt\}\sum_{t}\tilde{\phi}^{(t)}_ {\l,\l-\hat{\nu}}(u)
\phi^{(t)}_{\l-\hat{\nu}+\hat{\mu}, \l-\hat{\nu}}(u)\no\\
&&~~~~~~=\frac{\s(u+|\l-\hat{\nu}|
w+\frac{n-2}{2}w-\frac{n-1}{2})}{\s(u+|\l-\hat{\nu}|
w+\frac{n}{2}w-\frac{n-1}{2})} \lt\{
\prod_{k\neq\nu}\frac{\s(\l_{\nu k}w-w)} {\s(\l_{\nu
k}w)}\rt\}\d^{\nu}_{\mu}.\no\eea Finally, the resulting solution
to the dual RE by the isomorphism transformation (\ref{ISO-F})
from the diagonal solution to RE is given by \bea
\tilde{\K}(\l|u)^{\nu}_{\mu}\hspace{-0.2cm}=\hspace{-0.1cm}
\frac{\s(u+|\l-\hat{\nu}|
w+\frac{n-2}{2}w-\frac{n-1}{2})}{\s(u+|\l-\hat{\nu}|
w+\frac{n}{2}w-\frac{n-1}{2})}\hspace{-0.1cm} \lt\{
\prod_{k\neq\nu}\frac{\s(\l_{\nu k}w-w)} {\s(\l_{\nu
k}w)}\rt\}\hspace{-0.1cm}k(\l+\frac{n}{2}\bar{\e}|-u-\frac{nw}{2})_{\nu}
\d^{\nu}_{\mu}.\label{Solution-DRE}\eea Substituting the diagonal
solution of RE (\ref{Solution-RE}) into the above equation and
after redefining the boundary parameter $\bar{\xi}$ and the free
non-vanishing function $f(u,\l)$, one finds that the resulting
diagonal solution (\ref{Solution-DRE}) to the dual RE is exactly
the same as (\ref{Solution}).

\section{Conclusion and comments}
\label{Con} \setcounter{equation}{0}

By using the face-vertex correspondence (\ref{Face-vertex}) and
the isomorphism (\ref{ISO-V}) between the solutions to the RE and
its dual  for the $\Zb_n$ Belavin R-matrix, we construct the
isomorphism between the solutions to the RE and its dual for the
$A^{(1)}_{n-1}$ SOS R-matrix. By directly solving the equation, we
obtain a diagonal solution to the dual RE.  Our solution to the
dual RE can also be obtained through the isomorphism
transformation (\ref{ISO-F}) from the diagonal solution to RE
obtained in \cite{Bat96} by a special choice of the free parameter
$\l'$ (\ref{relation-2}).  Furthermore, the diagonal
$\tilde{\K}(\l|u)$ obtained in this paper enables us to
diagonalize the {\it double-row transfer matrices} of  the $\Zb_n$
Belavin model with open boundary condition described by the
diagonal $\K(\l|u)$ and the diagonal $\tilde{\K}(\l|u)$
\cite{Sas03}.

Alternatively in \cite{Qua01}, the very isomorphism with the
special choice of the free parameter $\l'$ (\ref{relation-2}) from
the diagonal solution of RE to the diagonal solution of the dual
RE was constructed by fusion procedure. However, our {\it
generic\/} isomorphism transformation (\ref{ISO-F}) gives a way to
construct a {\it non-diagonal\/} solution of the dual RE with
additional free parameters $\{\l'_i\}$.

\section*{Acknowledgements}
The support from  the Japan Society for the Promotion of Science
is gratefully acknowledged.


\end{document}